\documentclass{article}
\usepackage{amsmath}
\usepackage{amssymb}
\usepackage{graphicx}
\begin{document}

\begin{center}
\large
\textbf{A simple and quantum-mechanically motivated}
\\[0,1 cm]
\textbf{characterization of the formally real Jordan algebras}
\\[1,5 cm]
\normalsize
Gerd Niestegge
\\[0,3 cm]
\footnotesize
gerd.niestegge@web.de
\\[1,5 cm]
\end{center}
\normalsize

\begin{abstract}
Quantum theory's Hilbert space apparatus 
in its finite-dimen\-sion\-al version
is nearly reconstructed from four simple and quantum-mechanically 
motivated postulates for a quantum logic. 
The reconstruction process is not complete, since 
it excludes the two-dimensional Hilbert space and 
still includes the exceptional Jordan algebras, 
which are not part of the Hilbert space apparatus.
Options for physically meaningful 
potential generalizations of 
the apparatus are discussed.
\\[0,3 cm]
\textbf{Key Words:} Jordan algebra; quantum logic; transition probability; higher-order interference
\\[0,3 cm]
\textbf{PACS:} 03.65.Ta, 02.30.Tb
\newline
\textbf{MSC:} 17C55; 81P10; 81P16
\\[1,0 cm]
\end{abstract}

\noindent
\large
\textbf{1. Introduction}
\normalsize
\\[0,5 cm]
Quantum mechanics needs the vast mathematical Hilbert space formalism,
but does not provide a plausible derivation for it.
Attempts to derive this formalism, including its statistical interpretation,
from a small number of simple and plausible postulates 
have a long history. 
The focus changed from the algebraic and quantum logical approaches in earlier years
\cite{birkhoff-vN36, gunson1967algebraic, guz1981conditional, Keller1980, 
mackey1960lecture, piron1964axiomatique, segal1947postulates, sherman1956segal, 
soler1995characterization, varadarajan1968and1970, von1933algebraic} 
to the operational and information theoretic approaches,
often using the generalized probabilistic theories 
or sometimes category theory as their framework,
 in later years
\cite{barnum2019strongly, e19060253, 
barnum2014higher, chiribella2011informational, Dakic2011, de2012deriving, 
Fivel2011, Goyal2010, hardy2001quantum, Hoehn_Wever2017,
leifer2013towards, Ludwig:98375, masanes2011derivation, 
selby2018qth_from_diagrammatic, tull2018categorical, wetering2018, wilce2012four}.

The present paper resumes the quantum logical approach, considering an abstract model
of projective quantum measurement therein. A major postulate becomes the non-existence of 
third-order interference. Third-order interference and its absence in quantum mechanics
were discovered by Sorkin in 1994 \cite{sorkin1994quantum} and were not known in the earlier years.
This has become a matter of experimental and 
theoretical research in the recent past 
\cite{e19060253, barnum2014higher, dakic2014density, hickmann2011born, Lee-Selby2016, 
niestegge2013three, park2012three, sinha2010ruling, sollner2012testing}. 
\newpage
Selecting two further basic features of quantum mechanics as postulates, 
it is shown that, in the finite-dimensional case, 
the quantum logic can be represented as the system of idempotent elements 
in a formally real Jordan algebra;
this type of algebra was introduced and classified 
by J. von Neumann, P. Jordan and E. Wigner \cite{von1933algebraic}.

This paper reuses some of the author's 
earlier results \cite{nie2012AMP}. These results also apply in 
the infinite-dimensional case, but the restriction to finite dimensions 
in the present paper renders possible 
to sharpen the result 
(in \cite{nie2012AMP}, the quantum logic is only dense in, but not 
identical to the system of idempotent elements)
as well as 
to significantly simplify the 
postulates and to introduce them immediately 
into the framework of the quantum logic and its states. 
Some postulates needed in \cite{nie2012AMP} (power-associativity
and positivity of squares)
require the derivation of some additional structure (an algebra which is
neither commutative nor associative in the general case
and which will be used later to prove the major results in section 5)
before they can be defined; in the present paper, 
these postulates are derived from different and simpler ones.

Most recent operational and information theoretic approaches are also restricted 
to the finite-dimensional case. Moreover, they often include only
the simple (or irreducible) Jordan algebras, although some quantum-mechanical applications
(e.g., superselection)
require the use of non-simple algebras of observables. 
The approach of the present paper 
includes the non-simple case. Furthermore, it differs from all other papers 
in the usage of one special postulate: the existence and the uniqueness
of the conditional probabilities on the quantum logic (which 
turn out to become the abstract model of 
projective quantum measurement mentioned above).

The postulates are introduced in the next section.
After a brief outline of the formally real Jordan algebras in section 3,
some auxiliary results are derived from the postulates
in section 4. 
The main results are presented and proven in section 5.
Options for physically
reasonable potential generalizations of the mathematical apparatus of
quantum theory are discussed in the concluding section.
\\[0,5 cm]
\large
\textbf{2. The four postulates}
\normalsize
\\[0,5 cm]
The quantum logic of the usual quantum mechanics consists of the observables 
with the simple spectrum $\left\{ 0,1 \right\}$ 
(or, equivalently, of the self-adjoint projections or of the closed linear 
subspaces of the Hilbert space)
and forms an orthomodular lattice.
Originally, therefore, a quantum logic was mostly assumed to be an 
orthomodular lattice. However, there is no physical motivation 
for the existence of lattice operations
for propositions which are not compatible, and later a quantum logic 
was often assumed to be an orthomodular partially ordered set only.
So this is what is done here.
Different more general structures are conceivable and one was sometimes
used in the author's previous work; however, they involve more mathematical 
subtleties, which shall be avoided here.
\newpage
In this paper, a \textit{quantum logic} shall be an orthomodular partially ordered set $L$ 
with order relation $\leq$, 
smallest element $0$, largest element $\mathbb{I}$ and an orthocomplementation $'$.
This means that the following conditions are satisfied by the $p,q \in L$:
\begin{enumerate}
\item[(a)] $ q \leq p$ \textit{implies} $p' \leq q'$.
\item[(b)] $(p')' = p$.
\item[(c)] $p \leq q'$ \textit{implies that} $p \vee q$, \textit{the supremum of} $p$ \textit{and} $q$, \textit{exists}.
\item[(d)] $p \vee p' = \mathbb{I}$.
\item[(e)] Orthomodular law: $q \leq p$ \textit{implies} $p = q \vee (p \wedge q')$.
\end{enumerate}
Here, $p \wedge q$ denotes the infimum of $p$ and $q$, which exists iff $p' \vee p'$ exists.

The elements of the quantum logic are called \textit{propositions}. 
A proposition $e$ is called \textit{minimal} 
if there is no proposition $q$ with $q \leq e$ and $0 \neq q \neq e$.
The minimal propositions are also called \textit{atoms} in the common literature.
Two propositions $p$ and $q$ are \textit{orthogonal}, if $p \leq q'$ or, equivalently, $q \leq p'$;
in this case, $p \vee q$ shall be noted by $p + q$ in the following.
The interpretation of this mathematical terminology is as
follows: orthogonal events are exclusive, $p'$ is the negation of
$p$, and $p + q := p \vee q $ is the disjunction of the two exclusive events, $p$ and $q$.

It is not assumed either that the quantum logic $L$ is a lattice 
(in a lattice, $p \vee q$ and $p \wedge q$ exist for all lattice elements $p$ and $q$)
or that it satisfies 
the so-called covering property. Both play important roles in 
the early quantum logical approaches \cite{birkhoff-vN36, Keller1980, piron1964axiomatique,
soler1995characterization, varadarajan1968and1970}.

Let $V$ denote the linear space of the bounded real-valued functions on $L$, which are 
additive for orthogonal propositions. A \textit{state} allocates probability values to the propositions
and is an element $\mu \in V$ 
with $\mu(p) \geq 0$ for all $p \in L$ and $\mu(\mathbb{I})=1$.

Suppose a state $\mu$ and a proposition $p$ with $\mu(p) \neq 0$. 
If one wishes to extend the usual \textit{conditional probabilities} from standard probability theory
to this setting, the minimum requirement 
for the conditioned state $\mu_p$ of $\mu$ under $p$
is that
\begin{center}
$ \mu_p(q) = \frac{1}{\mu(p)} \mu(q) $ for $q \leq p$.
\end{center}
In the following, it shall be
assumed that such a state $\mu_p$ exists 
for each state $\mu$ and each proposition $p$ with $\mu(p) \neq 0$
and that it is unique (i.e. there is only one such state $\mu_p$).
\\[0,3 cm]
\textbf{Definition 2.1.} 
\itshape
In accordance with the usual case, the \emph{conditional probability} of a proposition $q$ 
under a further proposition $p$ in a state $\mu$ with $\mu(p) \neq 0$ is denoted by 
$$\mu(q|p) := \mu_p(q).$$
\normalfont

The assumption that unique conditional probabilities exist
does not hold for many quantum logics, but it does hold for the
Hilbert space quantum logic used in quantum mechanics 
(except for the two-dimensional case), and it turns out 
that the above transition from $\mu$ to the conditioned state then becomes identical 
to the state transition of the projective quantum measurement 
(L\"uders - von Neumann measurement; 
see \cite{niestegge2001non, niestegge2008approach}
and also remark 3.2 in section 3).
On the one hand, this 
provides a strong quantum-mechanical motivation 
for the above extension of conditional probability.
On the other hand, it is quite surprising that 
a natural extension of the classical conditional probability 
from the Boolean lattices to the quantum logics considered here
brings us to quantum measurement, 
the behavior of which is far away from any classical conceptions
and the source of many well-known interpretational and philosophical 
troubles with quantum mechanics.

That projective quantum measurement is the appropriate rule for 
conditionalizing probabilities in the non-Boolean quantum logic of quantum mechanics
has already been pointed out by J. Bub \cite{bub1977neumann} and others \cite{friedman1978quantum}.
Conditional probabilities, although not always defined in the same way as above,
were considered in the early attempts to derive the quantum-mechanical formalism 
\cite{guz1980conditional, guz1981conditional, pool1968baser}, 
but their existence and uniqueness was used as a postulate 
for the first time in 2001 in \cite{niestegge2001non}.
Further various ways of introducing 
conditional probabilities on quantum logics can be found in \cite{nanasiova1986conditional, 
nanasiova2000representation, nanasiova2015calculus, nanasiova2010maps, pulmannova1990quantum, redei1998quantum}.

In the case of a minimal proposition $e$, the state
$\mu_e : q \rightarrow \mu(q|e)$, $q \in L$, becomes independent of $\mu$ and
 is the unique state allocating the probability $1$ to the minimal proposition $e$. 
\\[0,3 cm]
\textbf{Definition 2.2.}
\itshape
Let $e \in L$ be a minimal proposition.
The \emph{state-independent} conditional probability of the propositions $q \in L$ 
under $e$ is denoted by $\mathbb{P}(q|e)$; this means that 
the following identity holds 
for all states $\mu$ with $\mu(e) \neq 0$ and all propositions $q \in L$\textnormal{:}
$$\mathbb{P}(q|e) = \mu(q|e).$$
\normalfont

The values of $ \mathbb{P}( \ | \ ) $ are intrinsic properties 
of the algebraic structure of the quantum logic 
and not primarily of its state space; they are invariant under the
algebraic automorphisms of the quantum logic.
Furthermore, $\mathbb{P}(q|e) = 1$ for any $q \in L$ with $e \leq q$, and 
$\mathbb{P}(q|e) = 0$ for any $q \in L$ with $e \leq q'$ (i.e., $e$ and $q$ are orthogonal).
Indeed, $0$ and $1$ are the only values
which $\mathbb{P}(q|e)$ can assume in a classical (i.e., Boolean)
logic, while any value in the unit interval 
is assumed in the quantum case.

The minimal propositions in the Hilbert space quantum logic are the lines in the Hilbert space, 
and the transition probability becomes $\mathbb{P}(e_2|e_1) = cos^{2} \theta$,
where $\theta$ is the angle between the two lines $e_1$ and $e_2$ 
\cite{niestegge2001non, niestegge2008approach}. We then have
$\mathbb{P}(e_2|e_1) = \mathbb{P}(e_1|e_2)$. This interesting symmetry
featured by the quantum-mechanical transition probabilities
will also become one of the postulates in this paper.
A similar axiom was used in many other studies, 
such as \cite{AS02, araki1980characterization, guz1981conditional}.

Moreover, the lines in the Hilbert space represent the quantum-mechanical pure states, 
and the mixed states can be expressed as combinations of orthogonal pure states, 
which motivates the assumption that the elements in $V$ are combinations of 
state-independent conditional probabilities generated by some orthogonal minimal propositions.
A similar assumption occurs in \cite{barnum2019strongly, barnum2014higher}; 
they do not use the state-independent conditional probabilities, but pure states instead,
and postulate the orthogonal decomposition only for the states and not for all elements of $V$.

A further feature of quantum mechanics is the impossibility of \emph{third-order interference}, 
which was discovered by Sorkin \cite{sorkin1994quantum}. 
Third-order interference is 
motivated by the consideration of three-slit experiments and the question of
whether they involve a new type of interference versus the common two-slit experiments. 
This can be expressed with the conditional
probabilities in the following way.
\\[0,3 cm]
\indent {$\mu(q|p_1+p_2+p_3) \mu(p_1+p_2+p_3)$
\\[0,3 cm]
\indent $-\mu(q|p_1+p_2) \mu(p_1+p_2)-\mu(q|p_1+p_3) \mu(p_1+p_3) - \mu(q|p_2+p_3) \mu(p_2+p_3)$
\\[0,3 cm]
\indent $+\mu(q|p_1) \mu(p_1)+\mu(q|p_2) \mu(p_2)+\mu(q|p_3) \mu(p_3)$,
\\[0,3 cm]
where $p_1,p_2,p_3$ are three orthogonal propositions, $q$ is a 
further proposition and $\mu$ is a state. Third-order interference means that 
there are states and propositions making this term non-zero.

The four postulates that the quantum logic $L$ shall satisfy can now be stated in the following way.
\itshape
\begin{enumerate}
\item[\emph{(A)}]
For every pair of propositions $p$ and $q$ in $L$ with $p\neq q$,
there is a state $\mu$ with $\mu(p) \neq \mu(q)$.
The (extended) conditional probabilities exist and are unique.
\item[\emph{(B)}]
The dimension of $V$ is finite, and each $\mu \in V$ can be written as
\begin{center}
$\mu(p) = \sum^{m}_{k=1} r_k \mathbb{P}(p|e_k)$, $p \in L$,
\end{center}
with some orthogonal minimal propositions $e_1,...,e_m$, 
real numbers $r_1,...,r_m$ and some positive integer $m$.
\item[\emph{(C)}]
For any two minimal propositions $e$ and $f$ in $L$, 
the transition probability satisfies the symmetry condition 
\begin{center}
$\mathbb{P}(f|e) = \mathbb{P}(e|f)$.
\end{center}
\item[\emph{(D)}]
Third-order interference does not occur.
\end{enumerate}
\normalfont
In \cite{barnum2014higher}, the non-existence of third-order interference
is combined with a certain strong symmetry postulate, which implies
the symmetry for the transition probabilities (C) considered here 
and which rules out the non-simple or reducible Jordan algebras
(except for the classical case). 
In \cite{barnum2019strongly}, it is shown that the non-existence of third-order interference
is a redundant postulate in \cite{barnum2014higher}. Regardless of the different frameworks,
the present paper takes another
approach: keeping the non-existence of third-order interference 
and using the weaker symmetry postulate (C), Jordan algebras 
can be derived in such a way that the non-simple or reducible ones 
are included.
\\[0,5 cm]
\large
\textbf{3. Jordan algebras}
\\[0,5 cm]
\normalsize
A \textit{Jordan algebra} is a linear space $A$ with a commutative bilinear product $\circ$ 
satisfying the identity $(x^{2} \circ y) \circ x = x^{2} \circ (y \circ x)$ for $x,y \in A$.
A Jordan algebra over the real numbers is called \textit{formally real}, if 
$x^{2}_{1} + ... + x^{2}_{m} = 0$ implies $x_1 = ... = x_m = 0$ for any $x_1,...,x_m \in A$
and any positive integer $m$. In the finite-dimensional case, the formally real Jordan algebras
coincide with the so-called JB-algebras and JBW-algebras \cite{AS02, hanche1984jordan}.

Each finite-dimensional formally real Jordan algebra possesses 
a multiplicative identity $\mathbb{I}$ and a natural order relation;
the system of its idempotent elements becomes a quantum logic with 
$p' := \mathbb{I} - p$.

A finite-dimensional formally real Jordan algebra decomposes 
into a direct sum of simple or irreducible subalgebras, 
which are either one-dimensional (the real numbers), spin factors 
(a spin factor is characterized by the conditions that all propositions in
the quantum logic formed by its idempotent elements are either minimal or 
identical to $0$ or $\mathbb{I}$
and that there are at least three different minimal propositions)
or can be represented 
as algebras of the Hermitian \textit{k}$\times$\textit{k}-matrices
over the real numbers, complex numbers, quaternions with $k = 3,4,5,...$ or 
over the octonions with $k=3$ only
\cite{hanche1984jordan, von1933algebraic}.
The product for the matrices $x,y$ is given by $x \circ y := (xy + yx)/2$.

Almost all finite-dimensional formally real Jordan algebras have a representation
as a Jordan subalgebra of a complex matrix algebra or, equivalently, 
the self-adjoint operators on a unitary space. Such a representation is not possible 
only for the so-called exceptional algebras; these are particularly 
the algebra that consists of Hermitian 3$\times$3-matrices over the octonions 
and all algebras that include this one as one of their irreducible subalgebras 
\cite{hanche1984jordan}.
\\[0,3 cm]
\textbf{Proposition 3.1.}
\itshape
The system of idempotent elements 
in a finite-dimensional formally real Jordan algebra $A$ 
yields a quantum logic satisfying the above postulates 
\normalfont
(A), (B), (C) \textit{and} (D), \textit{if none of the 
irreducible subalgebras of $A$ is a spin factor}.
\\[0,3 cm]
\textbf{Proof.}
Let $A$ be a finite-dimensional formally real Jordan algebra
such that none of its 
irreducible subalgebras is a spin factor.
In \cite{niestegge2001non}, it is shown 
that the unique conditional probabilities exist (postulate (A)) and 
have the shape 
\begin{equation*}
\tag{3.1}
\mu(q|p) = \frac{1}{\mu(p)}\mu(\left\{p,q,p\right\}),
\end{equation*}
with idempotent elements $p$ and $q$ in $A$, a state $\mu$ and $\mu(p) > 0$.
Here, $\left\{\ ,\ ,\ \right\}$ denotes the so-called Jordan triple product 
$\left\{x,y,z\right\} := x\circ(y\circ z) - y\circ(z\circ x) + z\circ(x\circ y)$ 
for $x,y,z \in A$.
Moreover, the unique extension of the state $\mu$ from the idempotent elements 
to the whole algebra $A$, which exists by an extension of Gleason's theorem 
to Jordan algebras \cite{bunce1985quantum, gleason1957measures}, is used
on the right-hand side of the above equation.
\newpage
The above special shape of the conditional probabilities implies that
third-order interference does not exist (postulate (D)) and, moreover,
that $\mathbb{P}(f|e) = r$ iff $\left\{e,f,e\right\} = re$
for minimal propositions $e$,$f$ and $r \in \mathbb{R}$.

$A$ possesses a natural real-valued \textit{trace}, defined by linearity 
and the condition that $trace(e) = 1$ holds for each minimal proposition $e$ \cite{AS02}.
It is identical to the trace inherited from the representation 
of $A$ as a direct sum of matrix algebras.
It satisfies
the identity $trace(\left\{p,q,p\right\}) = trace(p \circ q)$ 
for idempotent elements $p$ and $q$ in $A$,
which implies $\mathbb{P}(f|e) = trace(e \circ f) = trace(f \circ e) = \mathbb{P}(e|f)$ 
for minimal propositions $e$ and $f$; so we have (C).
Moreover, $A$ becomes a Hilbert space with the inner product
$\left\langle x|y \right\rangle := trace(x \circ y)$, $x,y \in A$. Hilbert spaces are self-dual
and the spectral theorem for the formally real Jordan algebras 
then yields postulate (B). 
\hfill $\square$
\\[0,3 cm]
\emph{Remark} 3.2.
To see the link to projective quantum measurement, 
which has already been mentioned above, 
note that the Jordan triple product 
$\left\{p,q,p\right\}$ coincides with the simple operator product $pqp$
in the case of the special Jordan product $x \circ y := (xy + yx)/2$ 
with Hilbert space operators $x$ and $y$. 
When the state $\mu$ is given by the statistical operator $a$ and $p,q$ are self-adjoint
projection operators on a Hilbert space,
equation (3.1) becomes 
\begin{equation*}
\tag{3.2}
\mu(q|p) = \frac{\mu(pqp)}{\mu(p)} = \frac{trace(apqp)}{trace(ap)} = \frac{trace(papq)}{trace(pap)} .
\end{equation*}
This shows that the state transition of projective quantum measurement (right-hand side of (3.2))
is the same as the probability conditionalization (left-hand side of (3.2)) in the case of the Hilbert space quantum logic.
\\[0,3 cm]
\textbf{Proposition 3.3.}
\itshape
If the irreducible subalgebras of a finite-dimensional
formally real Jordan algebra $A$ include a
spin factor, the quantum logic formed by its idempotent elements does not 
possess unique conditional probabilities.
\\[0,3 cm]
\normalfont
\textbf{Proof.}
Since unique conditional probabilities on $A$ enforce unique conditional probabilities
on each of its direct summands, it is sufficient to 
assume that $A$ itself is a spin factor.
Let $L$ be the system of its idempotent elements, $e$ and $f$ minimal propositions
that are neither identical nor orthogonal and $\mu$ a state with $\mu(e)=1$. 
Then define another state $\nu$ on $L$ that is identical to $\mu$ for all propositions
except for $f$ and $f'$. For these two propositions choose $\nu(f) \neq \mu(f)$ 
and $\nu(f') = 1 - \nu(f)$. 
Then $\nu(q) + \nu(q') = 1$ for $q = 0$, $q = \mathbb{I}$ and all 
minimal propositions $q$. Since $A$ is a spin factor, 
there are no further propositions and never 
more than two orthogonal non-zero propositions.
Therefore $\nu$ indeed becomes a state on $L$ and so there are
two different states $\mu \neq \nu$ with $\mu(e) = 1 = \nu(e)$. 
This contradicts the uniqueness of the conditional 
probability, which implies that there is only one single state on $L$
allocating the probability 1 to a given minimal proposition 
(this is the state $p \rightarrow \mathbb{P}(p|e)$). \hfill $\square$
\\[0,3 cm]
That a spin factor does not possess unique conditional probabilities is also 
an immediate consequence of the 
much more general lemma 3.3 in \cite{niestegge2001non}. 
\\[0,5 cm]
\large
\textbf{4. Auxiliaries}
\\[0,5 cm]
\normalsize
This section presents some simple auxiliary results, which will be used later 
and require only the first three postulates. The fourth one will be needed later.
\\[0,3 cm]
\textbf{Lemma 4.1.} 
\itshape
Let $L$ be a quantum logic satisfying the first three postulates \emph{(A)}, \emph{(B)} and \emph{(C)}.
\begin{enumerate}
\item[\emph{(i)}]
The cardinality of any family of pairwise orthogonal non-zero propositions in $L$ 
cannot exceed the dimension of $V$.
\item[\emph{(ii)}]
If $\sum^{l}_{i=1} e_i \leq \sum^{m}_{j=1} f_j$ with two families $e_1,...,e_l$ 
and $f_1,...,f_m$ of pairwise orthogonal minimal propositions, then $l \leq m$.
\item[\emph{(iii)}]
There is a minimal proposition $e$ in $L$ with $e\leq p$
for each proposition $p \neq 0$ in $L$.
\item[\emph{(iv)}] 
Each proposition $p\neq0$ in $L$ is the sum of a finite number of pairwise orthogonal
minimal propositions.
\item[\emph{(v)}] 
There is a unique state $\tau$ on $L$ with $\tau(e) = \tau(f)$ for all 
minimal propositions $e$ and $f$ in $L$. It is called the \emph{trace state} and has the shape
\begin{center}
$\tau(p) = \frac{1}{n} \sum^{n}_{k=1} \mathbb{P}(p|e_k)$,
\end{center}
for $p \in L$, where $e_1,...,e_n$ are any pairwise orthogonal minimal 
propositions with $\sum^{n}_{k=1} e_k = \mathbb{I}$.
\item[\emph{(vi)}] 
The conditional probability in the trace state $\tau$ satisfies the identity
\begin{center}
$ \tau(q|p) \tau(p) + \tau(q|p') \tau(p') = \tau(q)$,
\end{center}
for all propositions $q$ and $p\neq0$ in $L$.
\end{enumerate}
\normalfont
\textbf{Proof.} 
(i) Let $q_j$ be a family of pairwise orthogonal non-zero propositions in $L$.
For each single $j$, there is a state allocating a non-zero value to $q_j$ 
due to postulate (A), 
and this state can be conditioned under $q_j$; thus we get a state $\mu_j$
with $\mu_j(q_j) = 1$ for each $j$. Then $\mu_j(q_i) = 0$ for $i \neq j$. 
Therefore, the states $\mu_j$ are linearly independent and 
 the cardinality of both families $\mu_j$ and $q_j$ cannot exceed
the dimension of $V$.

(ii) Let $e_1,...,e_l$ and $f_1,...,f_m$ be two families of 
pairwise orthogonal minimal propositions with $\sum^{l}_{i=1} e_i \leq \sum^{m}_{j=1} f_j$ 
and define a state $\mu$ by
\begin{center}
$\mu(p) := \frac{1}{m} \sum^{m}_{j=1} \mathbb{P}(p|f_j)$ for $p \in L$.
\end{center}
Then $e_i \leq \sum^{m}_{j=1} f_j$, 
hence $\mathbb{P}(\sum^{m}_{j=1} f_j|e_i) = 1 $ for each $i$
and by postulate (C)
\begin{center}
$\mu(e_i) = \frac{1}{m} \sum^{m}_{j=1} \mathbb{P}(e_i|f_j) = \frac{1}{m} \sum^{m}_{j=1} \mathbb{P}(f_j|e_i) = \frac{1}{m} \mathbb{P}(\sum^{m}_{j=1} f_j|e_i) = \frac{1}{m}$
\end{center}
for $i = 1,...,l$. Therefore
\begin{center}
$1 = \mu(\sum^{m}_{j=1} f_j) \geq \mu(\sum^{l}_{i=1} e_i) = \sum^{l}_{i=1}\mu(e_i) = \frac{l}{m}$
\end{center}
and thus $l \leq m$.

(iii) Let $0 \neq p \in L$.
If there is no $q_1 \in L$ with $0 \neq q_1 \leq p$ and $q_1 \neq p$, 
the proposition $p$ itself is minimal and we are done. In the other case, consider 
$p \wedge {q_1}' \neq 0$. Either $p \wedge {q_1}'$ is minimal and we are done again or
there is a proposition $q_2 \in L$ with $0 \neq q_2 \leq p \wedge {q_1}'$ and $q_2 \neq p \wedge {q_1}'$.
This procedure is continued, but must stop after a finite number of steps by (i),
since the $q_1,q_2,...$ are pairwise orthogonal.

(iv) Let $0 \neq p \in L$. By (iii), there is a minimal proposition $e_1$ 
with $e_1 \leq p$. If $e_1 = p$, we are done. If not, consider $p \wedge {e_1}' \neq 0$
and again apply (ii) to get a minimal proposition $e_2$ 
with $e_2 \leq p \wedge {e_1}'$. If $e_2 = p \wedge {e_1}'$, we have $p = e_1 + e_2$.
If not, continue this procedure. By (i), it stops again after
a finite number of steps,
since the $e_1,e_2,...$ are pairwise orthogonal.

(v) By (iv), $\mathbb{I}$ is the sum of a finite number of pairwise orthogonal minimal
propositions $e_1,...,e_n$.
Define a state $\tau$ by
\begin{center}
$ \tau(q) := \frac{1}{n} \sum^{n}_{k=1} \mathbb{P}(q|e_k) $ for $q \in L $.
\end{center}
For any minimal proposition $f$ in $L$ we then have by postulate (C)
\begin{center}
$ \tau(f) = \frac{1}{n} \sum^{n}_{k=1} \mathbb{P}(f|e_k)
= \frac{1}{n} \sum^{n}_{k=1} \mathbb{P}(e_k|f) 
 = \frac{1}{n} \mathbb{P}(\sum^{n}_{k=1} e_k|f)
= \frac{1}{n} \mathbb{P}(\mathbb{I}|f) = \frac{1}{n} $.
\end{center}
Now assume that $\rho$ is a further state allocating identical values to the minimal
propositions. Then $1 = \rho(\mathbb{I}) = n \rho(e_1)$ and thus $\rho(f) = 1/n$ 
for all minimal propositions $f$. Since all propositions are sums of minimal propositions by (iv), $\rho$ must coincide with $\tau$.

(vi) Let $p \in L$. By (iv), $p=\sum^{m}_{k=1} e_k$ and $p'=\sum^{n}_{k=m+1} e_k$ with $n$ pairwise orthogonal minimal propositions $e_1,...,e_n$. Then $\mathbb{I}=p+p'=\sum^{n}_{k=1} e_k$. By (v), $\tau(q) = \frac{1}{n} \sum^{n}_{k=1} \mathbb{P}(q|e_k)$ for $q \in L$. It shall now be shown that the following two identities hold for $q \in L$:
\begin{equation*}
\tag{4.1}
\tau(q|p) = \frac{1}{m} \sum^{m}_{k=1} \mathbb{P}(q|e_k)
\end{equation*}
and
\begin{equation*}
\tag{4.2}
\tau(q|p') = \frac{1}{n-m} \sum^{m}_{k=1} \mathbb{P}(q|e_k).
\end{equation*}
If $q$ is a minimal proposition with $q \leq p$, the right-hand side of (4.1) yields, by postulate (C),
\begin{align*}
\frac{1}{m} \sum^{m}_{k=1} \mathbb{P}(q|e_k) & = \frac{1}{m} \sum^{m}_{k=1} \mathbb{P}(e_k|q)
 = \frac{1}{m} \mathbb{P}(\sum^{m}_{k=1} e_k|q) \\
& = \frac{1}{m} \mathbb{P}(p|q) = \frac{1}{m} = \frac{\tau(q)}{\tau(p)} = \tau(q|p).
\end{align*}
Note that $\tau(q) = 1/n$ and $\tau(p) = m/n$. Identity (4.1) then follows for all propositions $q \leq p$, since each such $q$ is a sum of minimal propositions, and the uniqueness of the conditional probability implies that it holds for all propositions $q$ in $L$.
Identity (4.2) follows in the same way. With (4.1) and (4.2), we then have, for any proposition $q$ in $L$,
\begin{align*}
\tau(q|p) \tau(p) + \tau(q|p') \tau(p') & = \tau(q|p) \frac{m}{n} + \tau(q|p') (1 - \frac{m}{n}) \\
 &= \frac{1}{n} \sum^{n}_{k=1} \mathbb{P}(q|e_k) = \tau(q).
\end{align*}
\hfill $\square$

\noindent
\large
\textbf{5. Results}
\\[0,5 cm]
\normalsize
In this section, some earlier results of \cite{nie2012AMP} will be needed; 
they are restated in the following two lemmas 
for the specific finite-dimensional case considered here.
\\[0,3 cm]
\textbf{Proposition 5.1.}
\itshape
Let $L$ be a quantum logic satisfying the three postulates 
\emph{(A)}, \emph{(B)} and \emph{(D)}. Then
\begin{enumerate}
\item[\emph{(i)}] $V$ is a base-norm space.
\item[\emph{(ii)}] $L$ can be embedded in the dual of $V$, which is an order-unit space denoted 
	by $A$, in such a way that $A$ is the linear hull of $L$,
	each element of $L$ is positive in $A$ and the order relations on $L$ and $A$ coincide.
\item[\emph{(iii)}] There is an idempotent positive linear map $U_p: A \rightarrow A$ 
	for each $p \in L \subseteq A$ such that the conditional probabilities have the shape 
	\begin{center}
	$\mu(q|p) \mu(q) = \mu (U_p q)$
	\end{center}
	 for any state $\mu$ on $L$ and any proposition $q$ in $L$
	\textnormal{(}note that $\mu \in V$ is identified with its canonical embedding 
	in the second dual $V^{**} = A^{*}$ here and in the following; 
	because of the finite dimension of $V$, we actually have $V = V^{**} = A^{*}$\textnormal{)}.
\footnote{
These $U_p$ projections bear some similarities to the so-called \textit{compressions} 
introduced and used by Alfsen and Shultz \cite{AS02}, but lack an important feature 
to become compressions: $U_p q = 0$ does not imply $U_{p'} q = q$ for $p,q \in L$.
}
\item[\emph{(iv)}] There is a bilinear operation $A \times A \rightarrow A$, $(a,b) \mapsto a \Box b$ 
	with 
	\begin{center}
	$p \Box q = \frac{1}{2} (q + U_p q - U_{p'} q) $,
	\end{center}
	 for $p,q \in L$.
\item[\emph{(v)}] $p^{2} := p \Box p = p$ for any $p \in L$ and $p \Box q = 0$ for any two orthogonal propositions $p,q \in L$.
\end{enumerate}
\normalfont
The first three items of this proposition follow from theorem 3.2 in \cite{nie2012AMP}; 
(iv) and (v) are shown in section 10 in \cite{nie2012AMP}.
Note that the so-called $\epsilon$-Hahn-Jordan decomposition property,
needed there, immediately follows from postulate (B).
Moreover, note that the orthomodular partially ordered set $L$ 
satisfies the slightly more general definition of a quantum logic used in \cite{nie2012AMP}.
Furthermore, because of the finite dimension of $V$, its dual $A$ is finite-dimensional as well;
this is why the different topologies 
used in \cite{nie2012AMP} need not be considered here, why all
linear subspaces become closed and why all linear maps are continuous.
\\[0,3 cm]
\textbf{Proposition 5.2.}
\itshape
Let $L$ be a quantum logic satisfying the three postulates 
\emph{(A)}, \emph{(B)} and \emph{(D)}. If the bilinear operation $\Box$ of proposition 5.1 
is power-associative and the squares $a^{2} := a \Box a$, $a \in A$, are 
positive elements in the order-unit space $A$, then the bilinear operation $\Box$
is commutative and $A$ is a formally real Jordan algebra.
\\[0,3 cm]
\normalfont
This immediately follows from theorem 11.1 in \cite{nie2012AMP}; in its proof, it is 
mentioned that $A$ becomes a so-called JBW/JB-algebra
and note that, in the finite-dimensional case considered here, the JBW/JB-algebras 
are identical to the formally real Jordan algebras.
\\[0,3 cm]
\textbf{Theorem 5.3.} \textit{A quantum logic $L$ which satisfies the four postulates \emph{(A)}, \emph{(B)}, \emph{(C)} and \emph{(D)} 
is identical to the quantum logic formed by the idempotent
elements of a finite-dimensional formally real Jordan algebra.}
\\[0,3 cm]
\textbf{Proof.}
Let $L$ be a quantum logic which satisfies the 
four postulates (A), (B), (C) and (D).
Lemma 4.1 (iv) implies that $\mathbb{I}$ is the sum of a finite number 
of orthogonal minimal propositions; the propositions vary, but lemma 4.1 (ii) implies 
that their number is fixed and is a typical characteristic of the quantum logic $L$.
In the following, let this be $n$. By lemma 4.1 (v), it follows that
$\tau(e) = 1/n$ for the trace state $\tau$ and all minimal propositions $e \in L$.

For any two minimal propositions $e$ and $f$ in $L$, we get, 
using proposition 5.1 (iii) and (iv) for the first equality,
lemma 4.1 (vi) for the second one, 
Definition 2.2 for the third one and finally postulate (C) for the last one,
\begin{align*}
\tau(f \Box e) & = \frac{1}{2}(\tau(f) + \tau(f|e) \tau(e) - \tau(f|e') \tau(e')) \\
& = \tau(f|e) \tau(e) = \mathbb{P} (f|e) \tau(e) 
 = \mathbb{P} (f|e)/n = \mathbb{P} (e|f)/n
\end{align*}
and thus $\tau(f \Box e) = \tau(e \Box f)$. 
Since the minimal propositions generate $A$, we get
\begin{center}
$\tau(a \Box b) = \tau(b \Box a)$ for all $a,b \in A$
\end{center}
and
\begin{center}
$\tau(e \Box p) = \tau(p \Box e) = \mathbb{P}(p|e)/n$
for any $p,e \in L$, with $e$ being minimal.
\end{center}
For any $x \in A$ define $\tau_x \in V$ by $\tau_x (p) := \tau (x \Box p)$, $p \in L$.
Owing to postulate (B), each $\mu \in V$ has this shape,
which can be seen in the following way:
there are orthogonal minimal propositions $e_1,...,e_m$, 
real numbers $r_1,...,r_m$ and some non-negative integer $m$ such that
$$\mu(p) = \sum^{m}_{k=1} r_k \mathbb{P}(p|e_k) = n \sum^{n}_{k=1} r_k \tau(e_k \Box p)$$
for $p \in L$; this means $\mu = \tau_a$ with
$$a = n \sum^{m}_{k=1} r_k e_k \in A.$$
Now consider $\tau_b \in V$ with any $b \in A$. 
Then there is some $a \in A$ with the above shape and 
$\tau_b = \tau_a$. For all $x \in A$, it follows that
\begin{align*}
\tau_x (b) &= \tau(x \Box b) = \tau(b \Box x) = \tau_b (x) \\
&= \tau_a(x) = \tau(a \Box x)= \tau(x \Box a)= \tau_x (a).
\end{align*}
Since each $\mu \in V$ has the shape $\mu = \tau_x$ with some $x \in A$,
this means $\mu(b) = \mu(a)$ for all $\mu \in V$ and finally $b = a$.
Therefore, each $b \in A$ is the linear combination of \emph{orthogonal} propositions from $L$
(i.e., $b$ has a spectral decomposition).

Now suppose $b = \sum^{m}_{k=1} s_k p_k \in A$ 
with orthogonal propositions $p_1,...,p_m \neq 0$ 
and real numbers $s_1,...,s_m$. Then by proposition 5.1 (v) and (ii)
\begin{center}
$b^{2} := b \Box b = \sum^{m}_{k=1} s_k^2 p_k \geq 0$
\end{center}
and, with $b^{l+1} := b \Box b^l$ for positive integers $l$,
\begin{center}
$b^{l} = \sum^{m}_{k=1} s_k^l p_k$.
\end{center}
This means that the square of each element in $A$ is a positive element in the order-unit space $A$
and that the bilinear operation $\Box$ is power-associative. 
By proposition 5.2, it is commutative and
$A$ becomes a formally real Jordan algebra.

Now assume $b =b^{2}$. Then
\begin{center}
$ \sum^{m}_{k=1} s_k p_k = \sum^{m}_{k=1} s_k^{2} p_k $.
\end{center}
The linear independence of $p_1,...,p_m$ implies $s_k = s_k^{2}$
and hence $s_k = 0$ or $s_k = 1$ for $k = 1,...,m$. 
Therefore, $b$ is the sum of orthogonal propositions in $L$ 
and thus itself an element of $L$. This means $L = \left\{ b\in A : b^2 = b \right\}$.
\hfill $\square$
\\[0,3 cm]
Combining theorem 5.3 with propositions 3.1 and 3.3 
immediately yields the following corollary.
\\[0,3 cm]
\textbf{Corollary 5.4.} 
\itshape
A quantum logic $L$ satisfies the four 
postulates \emph{(A)}, \emph{(B)}, \emph{(C)} and \emph{(D)} 
if and only if it is identical to the system of the idempotent
elements in a finite-dimensional formally real Jordan algebra whose 
irreducible subalgebras do not not include spin factors.
\normalfont
\newpage
\noindent
\large
\textbf{6. Conclusion}
\\[0,5 cm]
\normalsize
(a) \emph{Reconstruction of quantum mechanics} 
\\[0,3 cm]
Except for the so-called exceptional Jordan algebras,
each formally real Jordan algebra with finite dimension 
has a representation as a Jordan subalgebra of
the self-adjoint operators on a unitary space
(or, equivalently, self-adjoint quadratic matrices with complex entries). 
The above results thus 
come close to a derivation of the finite-dimensional case 
of quantum theory's mathematical apparatus from the four postulates (A), (B), (C) and (D).
Unfortunately, this is not a complete reconstruction of quantum mechanics,
since there remain two problems. 
The first one is that
the four postulates (A), (B), (C) and (D) rule out the spin factors and thus a certain part of
standard quantum mechanics: the two-dimensional Hilbert space and therefore also the qubit.
The second problem is that the four postulates (A), (B), (C) and (D) 
allow for the exceptional Jordan algebras,
which are not part of standard quantum mechanics.

Concerning the first problem, note that only the isolated single qubit is ruled out; 
the quantum-mechanical model of an $n$-qubit system is the algebra of 
$2^{n} \times 2^{n}$-matrices with
complex entries and its self-adjoint part 
(with the special Jordan product $x \circ y := (xy + yx)/2$) satisfies 
the postulates (A), (B), (C) and (D) for $n \neq 1$.

Some approaches to overcome the second problem and 
to find reasons for the necessity of the complex numbers 
in quantum mechanics are:
\\[0,3 cm]
\hspace*{3 cm}
(1) \emph{dynamical correspondence} \cite{AS02, nie2015dyn} 
\newline
\hspace*{3 cm}
(2) \emph{energy observable assignment} \cite{barnum2014higher}
\newline
\hspace*{3 cm}
(3) \emph{local tomography} \cite{barnum2019strongly, Bar_Wil2014locTom, barrett2007information,
chiribella2011informational, de2012deriving, Hardy_Foliable2011, masanes2011derivation}
\newline
\hspace*{3 cm}
(4) 3-\emph{ball property} \cite{AS02}
\newline
\hspace*{3 cm}
(5) \emph{orientations} \cite{AS02, AS01}.
\\[0,3 cm]
The first two approaches are closely related. Both concern
the mathematical model of the dynamical evolution of a single system and are
motivated by the quantum-mechanical feature that the dynamical group 
is generated by the Hamilton operator.

Local tomography means that
one can estimate the state of a composite system, 
composed of two or more subsystems,
by making measurements on its
subsystems (taking into account their correlations).
Local tomography could be phrased as an additional postulate directly 
in the quantum logical setting of the present paper, while this would be hard to 
do for the other approaches.
Though not explicitly stated in this way, the results in \cite{masanes2011derivation}
appear to imply that
local tomography distinguishes the complex case 
among the irreducible (simple) finite-dimensional formally real Jordan algebras 
\cite{barnum2019strongly},
but it seems to be unknown whether 
this holds generally when the reducible (non-simple) algebras are also considered.

The last two approaches involve mathematically interesting concepts, 
but a physical motivation is hard to find for them.
\\[0,3 cm]
(b) \emph{Potential generalizations}
\\[0,3 cm]
A different interesting problem is the question
of what opportunities there are for a physically meaningful generalization
of quantum theory's mathematical apparatus. The conditional probabilities 
represent an extension of projective quantum measurement to the quantum logical setting
and this may be a good reason to maintain postulate (A). 

Moreover, it has been shown in \cite{nie2012AMP} 
that the non-existence of third-order interference 
is responsible for the existence of the operation $\Box$ 
used in the proof of the above theorem,
and in \cite{nie2014GenQTh} 
a close link between this operation and the existence of continuous symmetries (Lie groups)
has been elaborated. A potential physically meaningful generalization of quantum mechanics 
should include continuous symmetries, and this may be an important reason 
to adhere to postulate (D) as well.

Therefore, only postulates (B) and (C) leave room for 
a potential generalization 
of quantum theory's mathematical apparatus.
The orthogonality of the minimal propositions $e_1,...,e_m$ in postulate (B)
is a very strong requirement and becomes one candidate 
to dispense with.
The symmetry of the transition probabilities is another candidate,
although it is often taken for granted. 
Some of the rare cases where asymmetric transition probabilities were considered in the past
are \cite{guz1980non, mielnik1969theory, mielnik1974generalized}.

The continuous symmetry groups of the finite-dimensional 
simple formally real Jordan algebras
cover all the non-exceptional compact simple Lie groups. 
Only five exceptional compact simple Lie groups remain; 
these are $G_2$, $F_4$, $E_6$, $E_7$, $E_8$. 
However, $F_4$ is also covered, since it is the Lie group of 
the exceptional Jordan algebra formed by the
Hermitian 3$\times$3-matrices over the octonions, 
and $G_2$ is the Lie group of the 
octonions themselves \cite{baez2002octonions}. 
Therefore, only $E_6$, $E_7$ and $E_8$ are not related 
to the symmetries of the formally real Jordan algebras 
and their associated quantum logics. 
Dropping postulate (B) or (C) or both may thus provide
opportunities to search for quantum logics 
with an $E_6$, $E_7$ or $E_8$ symmetry.
\\[0,3 cm]
(c) \emph{Soundness of plausibility considerations}
\\[0,3 cm]
Deriving the mathematical apparatus of quantum mechanics 
from a small number of mathematical features of this theory helps, 
on the one hand, to better understand why this apparatus is required and,
on the other hand, to study potential 
generalizations which might provide opportunities for future progress in theoretical physics. 
These features may be more or less physically motivated 
and those with less motivation become the first candidates to dispense with
in a potential generalization. 
However one must be careful. The implications of such a feature 
may not be obvious immediately 
and should be studied in the mathematical framework first. Moreover, physical 
plausibility depends on a scientist's personal background and belief. 
Could quantum theory ever have been detected on the basis of physical plausibility 
in the nineteenth century?
\newpage
At that time one would have insisted on 
commutative algebras for the observables 
and Boolean lattices for the propositions. 
Only later experimental evidence enforced 
the use of non-commutative algebras, resulting
in the non-classical quantum logics.

The conditional probabilities considered in this paper
are the only possible extension of the classical ones 
to quantum logics and, for the standard Hilbert space quantum logic,
it turns out that probability conditionalization 
is identical to the state transition of projective quantum measurement.
Its behavior is far away from any plausibility considerations 
from a classical point of view.

Hilbert space quantum theory includes second-order interference,
but excludes third- and higher-order interference.
At first glance, there are no plausible reasons why 
third- and higher-order interference should be ruled out,
while second-order interference is allowed. However, 
studying the implications in a generic mathematical framework
reveals a link to the existence of continuous symmetries,
which are quite important in physical theories.

Physically convincing reasons for the symmetry of the 
transition probabilities (C) and for postulate (B)
seem to be harder to find. Both require minimal propositions
which do not exist in some infinite-dimensional operator algebras
used in quantum statistical mechanics and in quantum field theory. 
The general validness of these two postulates in their current form
is thus ruled out by physical theories that already exist.

\bibliographystyle{abbrv}
\bibliography{Literatur}
\end{document}